 \documentclass[reprint,onecolumn,amsmath,amssymb,aip,jcp]{revtex4-1}

\usepackage{graphicx}% Include figure files
\usepackage{dcolumn}% Align table columns on decimal point
\usepackage{bm}% bold math
 \usepackage{float} 
\usepackage{fullpage}
\usepackage{color}
\usepackage{subcaption}
\usepackage{multirow}
\usepackage{amsmath}
\usepackage{amssymb}
\usepackage{amsthm}
\usepackage[colorlinks   = true, urlcolor     = cyan, linkcolor    = blue, citecolor   = red]{hyperref}
\usepackage{bm}
\usepackage{pict2e}
\usepackage{epstopdf}
\usepackage{comment}
\usepackage{epsfig}
\usepackage{url}
\usepackage[ruled,vlined]{algorithm2e}

\usepackage{cancel}
\usepackage{amsmath}
\usepackage{amssymb}
\usepackage{etoolbox}
\usepackage{graphicx}
 \usepackage[english]{babel}
\usepackage{mathrsfs} % Bilinear forms
\usepackage{enumitem}                                  % Fancy enumerate, e.g. a) b) c)

\usepackage{amsthm} % Theorems
\usepackage{siunitx}
\usepackage{ upgreek }

\begin{document}

%\preprint{APS/123-QED}

\title{Ab initio study of flexoelectricity in MXene  monolayers}

\author{Shashikant Kumar}
\affiliation{College of Engineering, Georgia Institute of Technology, Atlanta, GA 30332, USA}

\author{Zixi Zhang}
\affiliation{College of Engineering, Georgia Institute of Technology, Atlanta, GA 30332, USA}

\author{Phanish Suryanarayana}
\email{phanish.suryanarayana@ce.gatech.edu}
\affiliation{College of Engineering, Georgia Institute of Technology, Atlanta, GA 30332, USA}

%\date{\today}

\begin{abstract}
We investigate flexoelectricity in MXene monolayers from first principles. Specifically, we compute the transverse flexoelectric coefficients of 126 MXene monolayers along their two principal directions using Kohn-Sham density functional theory. The values span a wide range from 0.19\,$e$ to 1.3\,$e$ and are nearly isotropic with respect to bending direction. The transition metal is found to play a significant role in the flexoelectric response, with nitride-based MXenes consistently displaying larger coefficients than their carbide counterparts. Moreover, the coefficients increase with structural thickness, but when normalized by the bending modulus, which is also computed for all 126 monolayers, they exhibit the opposite trend.
\end{abstract}


\maketitle

%%%%%%%%%%%%%%%%%%%%%%%%%%%%%%%%%%%%%%%%%%%%%%%%%%%%%%%%%%%%%%
\section{Introduction}
The isolation of graphene \cite{geim2009graphene} has catalyzed rapid growth in the field of two-dimensional (2D) materials \cite{miro2014atlas, khan2020recent, butler2013progress}, leading to the discovery of numerous such materials. Among these are the MXenes \cite{lim2022fundamentals, lei2015recent, li2018functional, murali2022review}, which consist of transition metal carbides and nitrides. MXenes are described by the general formula $M_{n+1} X_nT_x$, where $n \in {1, 2, 3}$. Specifically, they consist of $(n+1)$ layers of transition metal atoms ($M$) alternating with $n$ layers of carbon or nitrogen ($X$), forming a layered structure. In addition, surface terminations ($T_x$) such as O, OH, or F may be present, bonded to the outermost metal layers. The first MXene to be experimentally  synthesized was Ti$_3$C$_2$T$_x$  in 2011 \cite{naguib2011two}. Since then, over 40 MXenes have been synthesized \cite{kumar2022methods}, and ab initio Kohn-Sham density functional theory (DFT) \cite{Hohenberg, Kohn1965} calculations have predicted the stability of hundreds more \cite{haastrup2018computational, zhou20192dmatpedia, nykiel2023high}.

MXenes exhibit a wide range of functional properties, including high electrical conductivity \cite{guo2018high, qiao2021electrical, lipatov2021high}, intrinsic hydrophilicity \cite{guan2020hydrophilicity, han2017preparation}, notable chemical reactivity \cite{wang2021stabilizing, yoo2023decoupling}, large negative zeta potential \cite{rozmyslowska2020engineering, jastrzkebska2017biological}, and efficient electromagnetic-wave absorption \cite{lu2024porous, song2020graphene}. These characteristics have enabled applications across diverse domains such as biomedicine \cite{soleymaniha2019promoting, zamhuri2021mxene} and electromagnetic interference shielding \cite{iqbal20202d, yun2020electromagnetic}. MXenes have also attracted growing interest for their mechanical \cite{yorulmaz2016vibrational, wyatt20212d} and electromechanical properties \cite{wu2020extraordinary, lipatov2020electrical, tan2019large, tan2021piezoelectricity, wang2022mxenes}, which include substantial in-plane stiffness \cite{hu2020quantifying}, electrically tunable Young’s modulus \cite{xu2022electrically}, and good mechanical strength \cite{shi2020interface, zhang2023scalable}. These capabilities make MXenes promising candidates for applications in energy harvesting, sensing \cite{pei2021ti3c2tx, ho2021sensing}, flexible electronics \cite{lyu2019large}, and nanomechanical devices. This motivates the accurate characterization of their mechanical and electromechanical response. 

The in-plane elastic constants for certain MXene monolayers have recently been determined through experiments \cite{lipatov2018elastic, lipatov2020electrical, lipatov2020electrical, rong2024elastic} as well as Kohn-Sham DFT studies \cite{zhang2018superior, bai2016dependence, fatima2023structural, zhang2018comprehensive}. In contrast, the response to bending deformations remains  unexplored, even though it is an important mode of deformation in 2D materials,  which are particularly susceptible to out-of-plane bending due to their  relatively low bending moduli. In particular, bending deformations can induce dipole moments through  flexoelectricity \cite{yudin2013fundamentals}, an electromechanical property intrinsic to semiconductors and insulators that describes the coupling between strain gradients and polarization. Although the flexoelectric effect is generally negligible for bulk systems, it becomes significant in 2D materials due to the possibility of extremely large strain gradients, especially  perpendicular to the sheet. However, no experimental data for the flexoelectric effect in MXenes is available, in part to the stringent curvature control and charge-detection sensitivity required. Although electronic structure calculations based on Kohn-Sham DFT can, in principle, furnish the flexoelectric coefficients, its cubic scaling with system size restricts practical simulations of bent structures to unrealistically large curvatures. Furthermore,  force fields that can capture the electronic origin of flexoelectricity, particularly in complex materials such as MXenes, are still lacking.

In this work, we investigate the flexoelectric effect in MXenes  from first principles by computing the transverse flexoelectric coefficients of 126 monolayers along their two principal directions using symmetry-adapted Kohn--Sham DFT. The calculated coefficients span a wide range, from 0.19\,$e$ to 1.3\,$e$, and exhibit near isotropy with respect to bending direction. The transition metal species is found to have a noticeable influence on the flexoelectric response, with nitride-based MXenes consistently yielding higher coefficients than their carbide analogs. Additionally, the flexoelectric coefficients increase with structural thickness; however, when normalized by the bending modulus, which is also computed for all 126 monolayers, the trend is reversed.

The remainder of this manuscript is organized as follows. In section~\ref{Sec:SysMeth}, we describe the MXene systems studied and the framework used for the symmetry-adapted Kohn–Sham DFT calculations. In section~\ref{Sec:Results}, we  present and discuss the results of these simulations. Finally, we provide concluding remarks  in section~\ref{Sec:Conclusions}.

%%%%%%%%%%%%%%%%%%%%%%%%%%%%%%%
%%%%%%%%%%%%%%%%%%%%%%%%%%%%%%%

\section{Systems and methods} \label{Sec:Methods}
We consider the following MXene monolayers: $M_{n+1}X_{n}T_x$, where $M \in \{ \text{Hf}, \text{Nb}, \text{Sc}, \text{Ta}, \text{Ti}, \text{V}, \text{Y}, \text{Zr} \}$, $X \in \{ \text{C}, \text{N} \}$, $T_x \in \{ \text{O}, \text{OH}, \text{F} \}$, and $n \in \{1, 2, 3\}$, yielding a total of $144 \times 2$ materials, with the factor of 2 accounting for the possibility of the  surface terminations $T_x$ being absent. Of these, 126 monolayers have been identified as dynamically and thermodynamically stable based on Kohn-Sham DFT calculations \cite{haastrup2018computational}, and thus only these are included in the analysis presented here. The complete list of materials studied this work can be found in Table~\ref{tab:data_flexo_bending}. 

The transversal flexoelectric coefficient for 2D materials can be defined as \cite{codony2021transversal}:
\begin{equation} \label{Eq:flexo_mu}
    \mu = \frac{\partial p_r}{\partial \kappa} \,,
\end{equation}
where $p_r$ is the radial polarization and $\kappa$ is the curvature associated with pure bending deformations. The radial polarization associated with a domain $\Omega$ can be determined from an electronic structure calculation such as Kohn-Sham DFT using the relation \cite{codony2021transversal}:
\begin{equation} \label{Eq:flexo_pr}
    p_r = \frac{1}{A}\int_{\Omega}(r-R_{\rm eff})\rho(\mathbf{x}) \, \mathrm{d\mathbf{x}} \,,
\end{equation}
where $A$ is the surface area of the deformed sheet in $\Omega$, $r$ is the radial coordinate corresponding to the spatial position $\mathbf{x}$, $R_{\rm eff}$ is the radial centroid of the ions in $\Omega$, and $\rho(\mathbf{x})$ is the electron density. The integrand in Eqn.~\ref{Eq:flexo_pr}, which can be interpreted as the radial dipole moment, is normalized by the area rather than the volume, unlike the standard definition of polarization in 3D bulk materials, since the thickness of 2D materials is not well defined.

\begin{figure}[h!]
	\centering
		\includegraphics[width=0.98\textwidth]{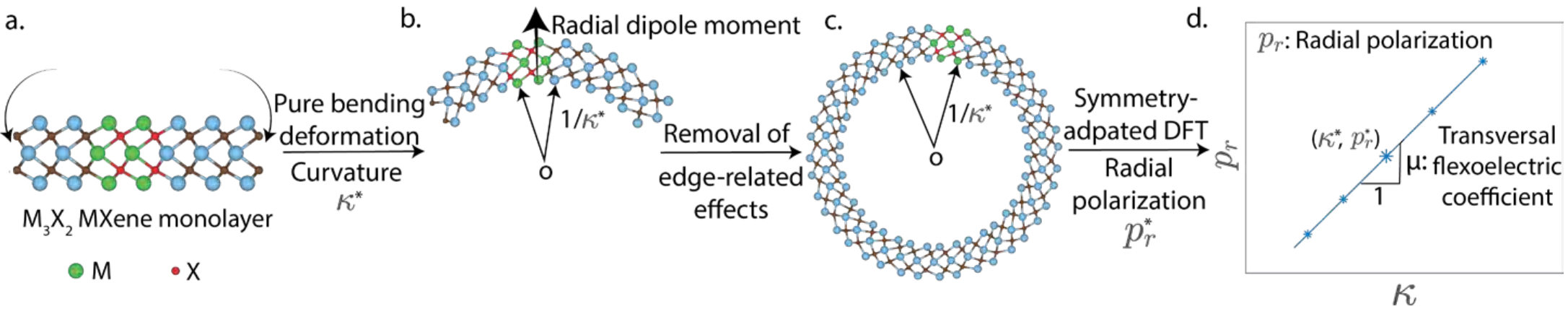}	
		\caption{Illustration of the framework employed to compute the transverse flexoelectric coefficient $\mu$ using cyclic symmetry-adapted Kohn-Sham DFT simulations. The atoms $M$ and $X$ in the unit cell of the MXene monolayer are colored green and red, respectively.}
		 \label{Fig:schematic}
\end{figure}

We compute the transversal flexoelectric coefficient $\mu$ using the procedure illustrated in Fig.~\ref{Fig:schematic}. Specifically, we employ a numerical approximation for the derivative of the radial polarization $p_r$  with respect to the curvature $\kappa$ in Eqn.~\ref{Eq:flexo_mu}. In particular, we compute the variation in $p_r$ with $\kappa$ and extract the slope from a curve that is fitted to the data. To compute $p_r$ at a specific  $\kappa$, edge-related effects are removed by periodically mapping the bent monolayer in the angular direction. The cyclic symmetry of the resultant nanotube structure is then exploited to perform Kohn–Sham DFT calculations using the Cyclix DFT \cite{sharma2021real} feature of the SPARC \cite{xu2021sparc, zhang2024sparc} electronic structure code. The number of atoms in the unit cell of the bent structure when using cyclic symmetry is the same as that in the unit cell for the flat sheet configuration when using translational/periodic symmetry, making the calculations particularly efficient.  
 
In all Kohn-Sham DFT simulations, we employ the Perdew\textendash Burke\textendash Ernzerhof (PBE)  \cite{perdew1986accurate} variant of the generalized gradient approximation (GGA) for the exchange-correlation functional and optimized norm-conserving Vanderbilt (ONCV) pseudopotentials \cite{hamann2013optimized} from the SPMS collection \cite{shojaei2023soft}. All numerical parameters, including grid spacing, $k$-point sampling for Brillouin zone integration in the cyclic and axial directions, vacuum spacing in the radial direction, and structural (cell and atom) relaxation tolerances, are selected to ensure that the computed flexoelectric coefficients are accurate to within 0.01$e$, as verified through convergence studies.

%%%%%%%%%%%%%%%%%%%%%%%%%%%%%%%%%%%%%%%%%%%%%%%%%%%%%%%%%%%%%%%%%%%%%%%%%%%%%%%%%%%%%%%%%%%%%%%%%%%%%%%%%%%%%%%%%%%%%%%%%%%%%%%%%%%%%%%%%%%%%%%%
\section{Results and discussion} \label{Sec:Results}
We calculate the transversal flexoelectric coefficient $\mu$ for the 126 MXene monolayers along their principal directions, i.e., armchair and zigzag, using the symmetry-adapted Kohn-Sham DFT framework described in the previous section.  We consider bending curvatures in the range $0.15 < \kappa < 0.25$ nm$^{-1}$, representative of those encountered in experiments \cite{lindahl2012determination,  han2019ultrasoft,  wang2019bending}.  In this regime, we have found the radial polarization to vary linearly with the curvature for all the MXenes, indicating linear response. We now discuss the results obtained, with the corresponding data provided in Table~\ref{tab:data_flexo_bending}.

%%%%%%%%%%%%%%%
\begin{table}[h!]
\centering
\caption{Flexoelectric coefficient $\mu$ and bending modulus $D$ for the MXene monolayers. Subscripts 1 and 2 denote values along the zigzag and armchair directions, respectively. The units of $\mu$ and $D$ are $e$ and eV, respectively.}
\resizebox{\textwidth}{!}{%
\begin{tabular}{lcccc|lcccc|lcccc}
\hline
MXene & $\mu_1$ & $\mu_2$  & $D_1$  & $D_2$  & MXene & $\mu_1$  & $\mu_2$ & $D_1$ & $D_2$ & MXene & $\mu_1$ & $\mu_2$ & $D_1$ & $D_2$ \\
\hline
Hf$_{4}$C$_{3}$ & 1.09 & 1.13 & 150.6 & 150.2 & Sc$_{4}$C$_{3}$O$_{2}$ & 0.40 & 0.45 & 190.4 & 179.9 & Y$_{2}$C & 0.36 & 0.35 & 5.850 & 5.840 \\
Hf$_{2}$C & 0.51 & 0.52 & 9.500 & 9.46 & Sc$_{4}$F$_{2}$N$_{3}$ & 0.83 & 0.81 & 177.7 & 172.6 & Y$_{2}$CF$_{2}$ & 0.38 & 0.37 & 14.10 & 14.47 \\
Hf$_{2}$CF$_{2}$ & 0.57 & 0.57 & 14.92 & 13.83 & Sc$_{4}$H$_{2}$N$_{3}$O$_{2}$ & 0.70 & 0.74 & 196.1 & 190.4 & Y$_{2}$CH$_{2}$O$_{2}$ & 0.48 & 0.44 & 16.75 & 16.60 \\
Hf$_{2}$CH$_{2}$O$_{2}$ & 0.47 & 0.50 & 19.01 & 17.82 & Sc$_{4}$N$_{3}$ & 0.85 & 0.83 & 111.5 & 108.4 & Y$_{2}$CO$_{2}$ & 0.28 & 0.31 & 36.75 & 36.86 \\
Hf$_{2}$CO$_{2}$ & 0.46 & 0.47 & 1.430 & 35.52 & Sc$_{4}$N$_{3}$O$_{2}$ & 0.31 & 0.42 & 214.7 & 205.2 & Y$_{2}$F$_{2}$N & 0.39 & 0.39 & 14.22 & 14.49 \\
Hf$_{3}$N$_{2}$O$_{2}$ & 0.75 & 0.76 & 150.5 & 139.8 & Ta$_{2}$CO$_{2}$ & 0.67 & 0.66 & 29.18 & 29.42 & Y$_{2}$H$_{2}$NO$_{2}$ & 0.49 & 0.51 & 15.56 & 15.95 \\
Hf$_{2}$H$_{2}$NO$_{2}$ & 0.66 & 0.64 & 14.50 & 13.08 & Ta$_{2}$N & 0.63 & 0.60 & 9.420 & 9.210 & Y$_{2}$N & 0.48 & 0.47 & 8.100 & 8.060 \\
Hf$_{2}$N & 0.58 & 0.60 & 12.05 & 12.08 & Ta$_{3}$C$_{2}$ & 0.93 & 0.85 & 50.63 & 49.59 & Y$_{3}$C$_{2}$ & 0.48 & 0.46 & 32.69 & 31.72 \\
Hf$_{2}$NO$_{2}$ & 0.48 & 0.54 & 33.91 & 31.77 & Ta$_{3}$C$_{2}$O$_{2}$ & 0.92 & 0.89 & 116.4 & 117.8 & Y$_{3}$C$_{2}$F$_{2}$ & 0.46 & 0.49 & 59.54 & 60.45 \\
Hf$_{3}$C$_{2}$ & 0.78 & 0.78 & 49.84 & 49.49 & Ta$_{3}$N$_{2}$ & 0.99 & 1.09 & 90.96 & 89.69 & Y$_{3}$C$_{2}$H$_{2}$O$_{2}$ & 0.76 & 0.76 & 64.81 & 66.48 \\
Hf$_{3}$C$_{2}$F$_{2}$ & 0.90 & 0.90 & 74.31 & 70.95 & Ta$_{4}$C$_{3}$ & 1.30 & 1.25 & 184.4 & 180.5 & Y$_{3}$C$_{3}$O$_{2}$ & 0.34 & 0.33 & 41.36 & 41.16 \\
Hf$_{3}$C$_{2}$H$_{2}$O$_{2}$ & 0.63 & 0.62 & 89.36 & 87.39 & Ta$_{4}$C$_{3}$O$_{2}$ & 1.22 & 1.13 & 290.4 & 281.8 & Y$_{3}$F$_{2}$N$_{2}$ & 0.51 & 0.60 & 72.44 & 71.33 \\
Hf$_{3}$C$_{2}$O$_{2}$ & 0.60 & 0.60 & 134.4 & 123.9 & Ti$_{2}$CF$_{2}$ & 0.56 & 0.56 & 13.80 & 14.21 & Y$_{3}$H$_{2}$N$_{2}$O$_{2}$ & 0.72 & 0.69 & 77.94 & 78.68 \\
Hf$_{3}$N$_{2}$ & 1.03 & 0.98 & 57.14 & 56.08 & Ti$_{2}$CH$_{2}$O$_{2}$ & 0.28 & 0.32 & 20.04 & 19.82 & Y$_{3}$N$_{2}$ & 0.65 & 0.64 & 40.27 & 39.64 \\
Hf$_{4}$C$_{3}$F$_{2}$ & 1.23 & 1.12 & 195.5 & 176.3 & Ti$_{2}$CO$_{2}$ & 0.53 & 0.53 & 31.95 & 31.21 & Y$_{3}$N$_{2}$O$_{2}$ & 0.22 & 0.27 & 93.23 & 90.01 \\
Hf$_{4}$C$_{3}$O$_{2}$ & 0.73 & 0.75 & 162.3 & 157.1 & Ti$_{2}$F$_{2}$N & 0.59 & 0.59 & 12.01 & 11.85 & Y$_{4}$C$_{3}$ & 0.32 & 0.30 & 81.41 & 78.76 \\
Hf$_{4}$N$_{3}$ & 1.24 & 1.23 & 160.5 & 158.0 & Ti$_{2}$H$_{2}$NO$_{2}$ & 0.37 & 0.37 & 18.63 & 17.79 & Y$_{4}$C$_{3}$F$_{2}$ & 0.56 & 0.61 & 157.6 & 147.5 \\
Hf$_{4}$N$_{3}$O$_{2}$ & 0.82 & 0.91 & 289.0 & 285.9 & Ti$_{2}$N & 0.45 & 0.45 & 7.760 & 7.600 & Y$_{4}$C$_{3}$H$_{2}$O$_{2}$ & 0.97 & 0.98 & 163.2 & 166.8 \\
Nb$_{2}$CO$_{2}$ & 0.67 & 0.65 & 28.10 & 27.66 & Ti$_{3}$C$_{2}$ & 0.57 & 0.52 & 34.57 & 34.69 & Y$_{4}$C$_{3}$O$_{2}$ & 0.24 & 0.20 & 198.4 & 191.8 \\
Nb$_{2}$N & 0.61 & 0.59 & 6.900 & 6.970 & Ti$_{3}$C$_{2}$F$_{2}$ & 0.88 & 0.84 & 66.22 & 63.78 & Y$_{4}$F$_{2}$N$_{3}$ & 0.66 & 0.62 & 179.2 & 173.1 \\
Nb$_{3}$C$_{2}$ & 0.78 & 0.78 & 39.48 & 38.94 & Ti$_{3}$C$_{2}$H$_{2}$O$_{2}$ & 0.39 & 0.39 & 82.07 & 79.38 & Y$_{4}$H$_{2}$N$_{3}$O$_{2}$ & 0.98 & 0.99 & 188.1 & 182.2 \\
Nb$_{3}$C$_{2}$O$_{2}$ & 0.92 & 0.91 & 110.1 & 112.6 & Ti$_{3}$C$_{2}$O$_{2}$ & 0.61 & 0.64 & 106.7 & 104.9 & Y$_{4}$N$_{3}$ & 0.96 & 0.99 & 116.8 & 113.4 \\
Nb$_{3}$N$_{2}$ & 1.03 & 0.99 & 76.82 & 76.27 & Ti$_{3}$N$_{2}$ & 0.67 & 0.68 & 33.48 & 33.54 & Y$_{4}$N$_{3}$O$_{2}$ & 0.20 & 0.17 & 216.7 & 204.1 \\
Nb$_{4}$C$_{3}$ & 1.15 & 1.13 & 153.3 & 151.4 & Ti$_{3}$N$_{2}$O$_{2}$ & 0.78 & 0.70 & 122.5 & 115.3 & Zr$_{2}$CF$_{2}$ & 0.56 & 0.55 & 15.19 & 14.30 \\
Nb$_{4}$C$_{3}$O$_{2}$ & 1.22 & 1.08 & 274.1 & 255.6 & Ti$_{4}$C$_{3}$ & 0.76 & 0.71 & 114.3 & 111.9 & Zr$_{2}$CH$_{2}$O$_{2}$ & 0.41 & 0.40 & 18.01 & 17.53 \\
Sc$_{2}$CF$_{2}$ & 0.39 & 0.39 & 15.01 & 14.49 & Ti$_{4}$C$_{3}$F$_{2}$ & 1.15 & 1.10 & 178.0 & 167.3 & Zr$_{2}$CO$_{2}$ & 0.46 & 0.51 & 33.56 & 33.10 \\
Sc$_{2}$CH$_{2}$O$_{2}$ & 0.49 & 0.47 & 18.43 & 17.79 & Ti$_{4}$C$_{3}$H$_{2}$O$_{2}$ & 0.49 & 0.45 & 202.6 & 194.4 & Zr$_{2}$N & 0.53 & 0.53 & 10.47 & 10.32 \\
Sc$_{2}$N & 0.44 & 0.45 & 7.670 & 7.300 & Ti$_{4}$C$_{3}$O$_{2}$ & 0.78 & 0.85 & 266.1 & 251.3 & Zr$_{2}$F$_{2}$N & 0.56 & 0.56 & 10.91 & 11.07 \\
Sc$_{2}$C & 0.29 & 0.31 & 4.150 & 4.200 & Ti$_{4}$H$_{2}$N$_{3}$O$_{2}$ & 0.81 & 0.80 & 175.5 & 172.6 & Zr$_{2}$H$_{2}$NO$_{2}$ & 0.49 & 0.52 & 14.63 & 14.55 \\
Sc$_{2}$CO$_{2}$ & 0.29 & 0.34 & 40.46 & 38.71 & Ti$_{4}$N$_{3}$ & 0.97 & 1.00 & 134.2 & 131.5 & Zr$_{2}$NO$_{2}$ & 0.46 & 0.53 & 30.78 & 31.26 \\
Sc$_{2}$F$_{2}$N & 0.45 & 0.44 & 14.62 & 14.00 & Ti$_{4}$N$_{3}$O$_{2}$ & 0.88 & 0.89 & 237.0 & 213.5 & Zr$_{3}$C$_{2}$ & 0.72 & 0.65 & 45.04 & 44.11 \\
Sc$_{2}$H$_{2}$NO$_{2}$ & 0.40 & 0.37 & 19.27 & 18.59 & V$_{2}$CF$_{2}$ & 0.58 & 0.63 & 12.44 & 12.49 & Zr$_{3}$C$_{2}$H$_{2}$O$_{2}$ & 0.55 & 0.57 & 84.33 & 83.66 \\
Sc$_{2}$NO$_{2}$ & 0.39 & 0.38 & 28.22 & 28.74 & V$_{2}$CH$_{2}$O$_{2}$ & 0.29 & 0.25 & 17.69 & 17.09 & Zr$_{3}$C$_{2}$O$_{2}$ & 0.60 & 0.56 & 121.8 & 114.3 \\
Sc$_{3}$C$_{2}$ & 0.44 & 0.36 & 30.28 & 29.80 & V$_{2}$CO$_{2}$ & 0.6 & 0.63 & 24.12 & 24.29 & Zr$_{3}$N$_{2}$ & 0.93 & 0.9 & 42.65 & 42.96 \\
Sc$_{3}$C$_{2}$H$_{2}$O$_{2}$ & 0.68 & 0.66 & 57.65 & 58.63 & V$_{2}$N & 0.47 & 0.53 & 3.780 & 3.630 & Zr$_{3}$N$_{2}$O$_{2}$ & 0.68 & 0.65 & 132.6 & 131.2 \\
Sc$_{3}$F$_{2}$N$_{2}$ & 0.39 & 0.39 & 71.47 & 71.10 & V$_{3}$C$_{2}$ & 0.61 & 0.63 & 33.93 & 32.83 & Zr$_{4}$C$_{3}$ & 0.93 & 0.87 & 134.6 & 131.9 \\
Sc$_{3}$H$_{2}$N$_{2}$O$_{2}$ & 0.43 & 0.47 & 84.83 & 82.03 & V$_{3}$C$_{2}$O$_{2}$ & 0.84 & 0.77 & 61.19 & 63.89 & Zr$_{4}$C$_{3}$F$_{2}$ & 1.09 & 1.02 & 192.6 & 182.9 \\
Sc$_{3}$N$_{2}$ & 0.54 & 0.51 & 33.98 & 34.09 & V$_{3}$N$_{2}$ & 0.75 & 0.74 & 34.29 & 33.71 & Zr$_{4}$C$_{3}$H$_{2}$O$_{2}$ & 0.74 & 0.75 & 209.5 & 203.2 \\
Sc$_{3}$N$_{2}$O$_{2}$ & 0.36 & 0.36 & 90.81 & 88.17 & V$_{3}$N$_{2}$O$_{2}$ & 0.76 & 0.75 & 85.86 & 102.3 & Zr$_{4}$C$_{3}$O$_{2}$ & 0.69 & 0.82 & 293.0 & 278.0 \\
Sc$_{4}$C$_{3}$ & 0.39 & 0.32 & 75.40 & 72.93 & V$_{4}$C$_{3}$ & 0.86 & 0.87 & 118.5 & 116.8 & Zr$_{4}$H$_{2}$N$_{3}$O$_{2}$ & 1.16 & 1.04 & 160.7 & 146.0 \\
Sc$_{4}$C$_{3}$F$_{2}$ & 0.79 & 0.80 & 158.0 & 147.3 & V$_{4}$C$_{3}$O$_{2}$ & 1.16 & 1.01 & 171.1 & 208.1 & Zr$_{4}$N$_{3}$ & 1.13 & 1.13 & 145.3 & 143.1 \\
Sc$_{4}$C$_{3}$H$_{2}$O$_{2}$ & 1.01 & 1.06 & 168.7 & 162.9 & V$_{4}$N$_{3}$ & 0.89 & 0.87 & 101.8 & 102.9 & Zr$_{4}$N$_{3}$O$_{2}$ & 0.77 & 0.82 & 281.5 & 265.8 \\
\hline
\end{tabular}%
}%
\label{tab:data_flexo_bending}
\end{table}
%%%%%%%%%%%%%%%%

The flexoelectric coefficient values are similar in the armchair and zigzag directions, with an average difference of 1.5\% and a maximum difference of 11\%, indicating minor directional anisotropy. This is consistent with the Kohn-Sham DFT results in literature for other atomic monolayers with a honeycomb structure, e.g., graphene \cite{kumar2021flexoelectricity, codony2021transversal}.  Interestingly, the flexoelectric coefficient has been found to be nearly isotropic even for rectangular monolayers \cite{Kumar2023Onthe}.  Given the nearly isotropic response,  the average of the flexoelectric coefficient values along the armchair and zigzag directions will be used in the subsequent analysis and discussion. 

The  transversal flexoelectric coefficient  values span a wide range, from $\mu = 0.19\,e$ for Y$_4$N$_3$O$_2$ to $\mu = 1.3\,e$ for Ta$_4$C$_3$.  In comparison, the flexoelectric coefficient of graphene from Kohn-Sham DFT calculations has been reported to be $\mu = 0.22\,e$ \cite{kumar2021flexoelectricity, codony2021transversal}.  The significantly larger value for Ta$_4$C$_3$ relative to graphene can be attributed in part to its larger thickness,  which allows for larger flexoelectric coefficients, given that the radial dipole moment is normalized by the area rather than the volume for 2D materials.  The smaller flexoelectric coefficient of Y$_4$N$_3$O$_2$ compared to graphene, despite its significantly larger thickness, suggests that the flexoelectric effect is nearly absent in this monolayer.

\begin{figure}[h!]
	\centering
		\includegraphics[width=0.85\textwidth]{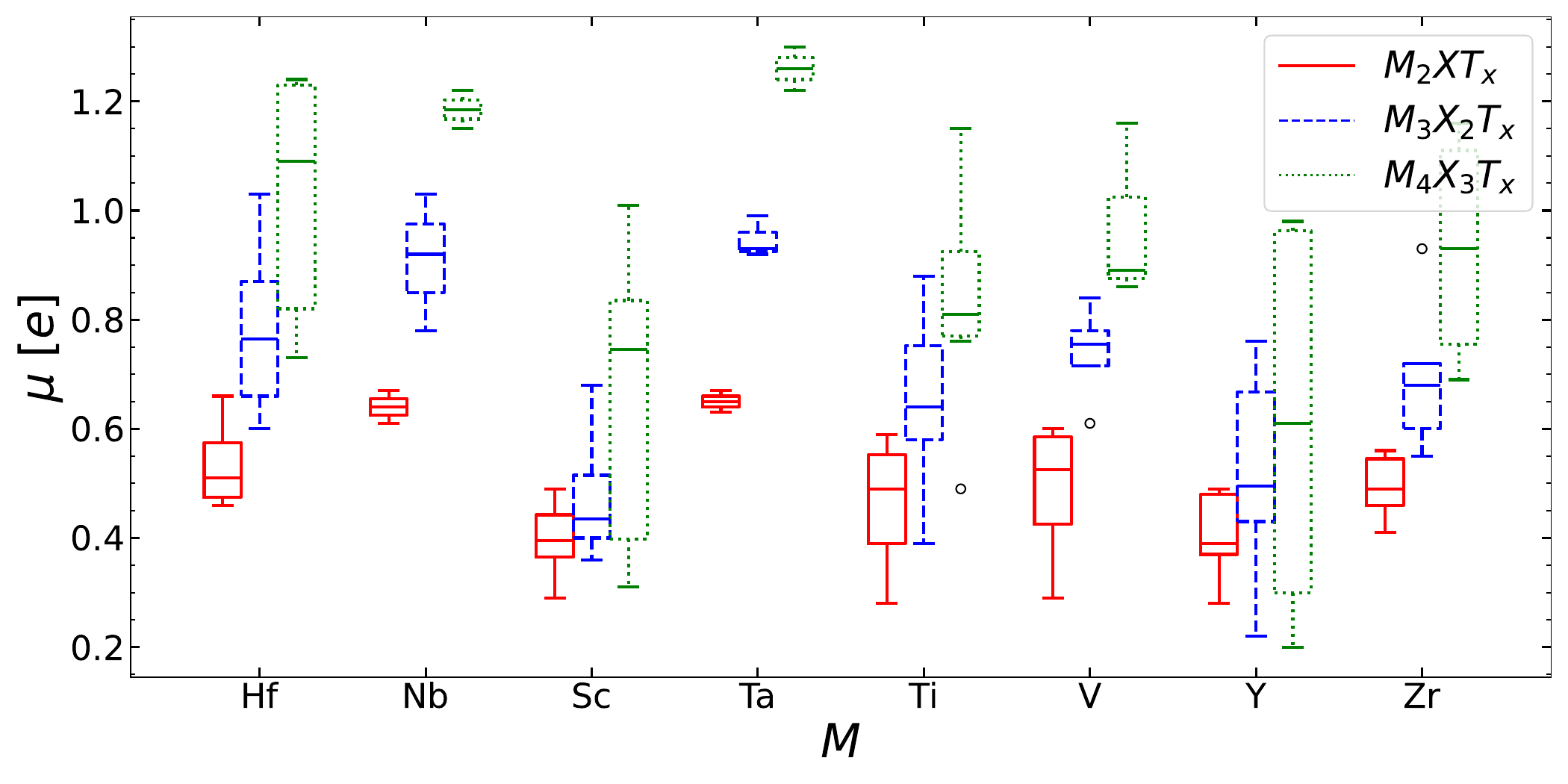}	
		\caption{Flexoelectric coefficient $\mu$ for the MXene monolayers, averaged over the armchair and zigzag directions, grouped by the transition metal  $M$.}
		 \label{Fig:flexobox}
\end{figure}

In Fig.~\ref{Fig:flexobox}, we present the flexoelectric coefficient values grouped by the transition metal $M$. The identity of the metal $M$ has a noticeable effect on the coefficient: Sc-based MXenes exhibit the smallest values, while Ta-based MXenes exhibit the largest. Averaged over different compositions, the  flexoelectric coefficients follows the trend: Sc $<$ Y $<$ Ti $<$ Zr $<$ V  $<$ Hf $<$ Nb $<$ Ta. This ordering remains largely consistent across variations in $X$ and surface terminations $T_{\rm x}$, underscoring the central role of the transition metal. Moreover, the trend suggests that the flexoelectric coefficient tends to increase with both the period and group number of the transition metal, likely due to the increased atomic size and electronic polarizability.

The results in Fig.~\ref{Fig:flexobox} also reveal a clear trend in the flexoelectric coefficients: $M_2XT_x < M_3X_2T_x < M_4 X_3 T_x$. This trend can be attributed to the increasing thickness of the MXene layers with larger $n$, which enhances the dipolar response to strain gradients. In particular, as discussed above, the radial dipole moment in 2D materials is normalized by area rather than volume, so systems with greater thickness tend to exhibit larger flexoelectric coefficients. This is consistent with previous Kohn-Sham DFT results reported in the literature \cite{kumar2021flexoelectricity}, where transition metal dichalcogenide/trichalcogenide monolayers were found to exhibit significantly larger flexoelectric coefficients compared to monolayers consisting of a single atomic layer, such as graphene.

\begin{figure}[h!]
	\centering
		\includegraphics[width=0.50\textwidth]{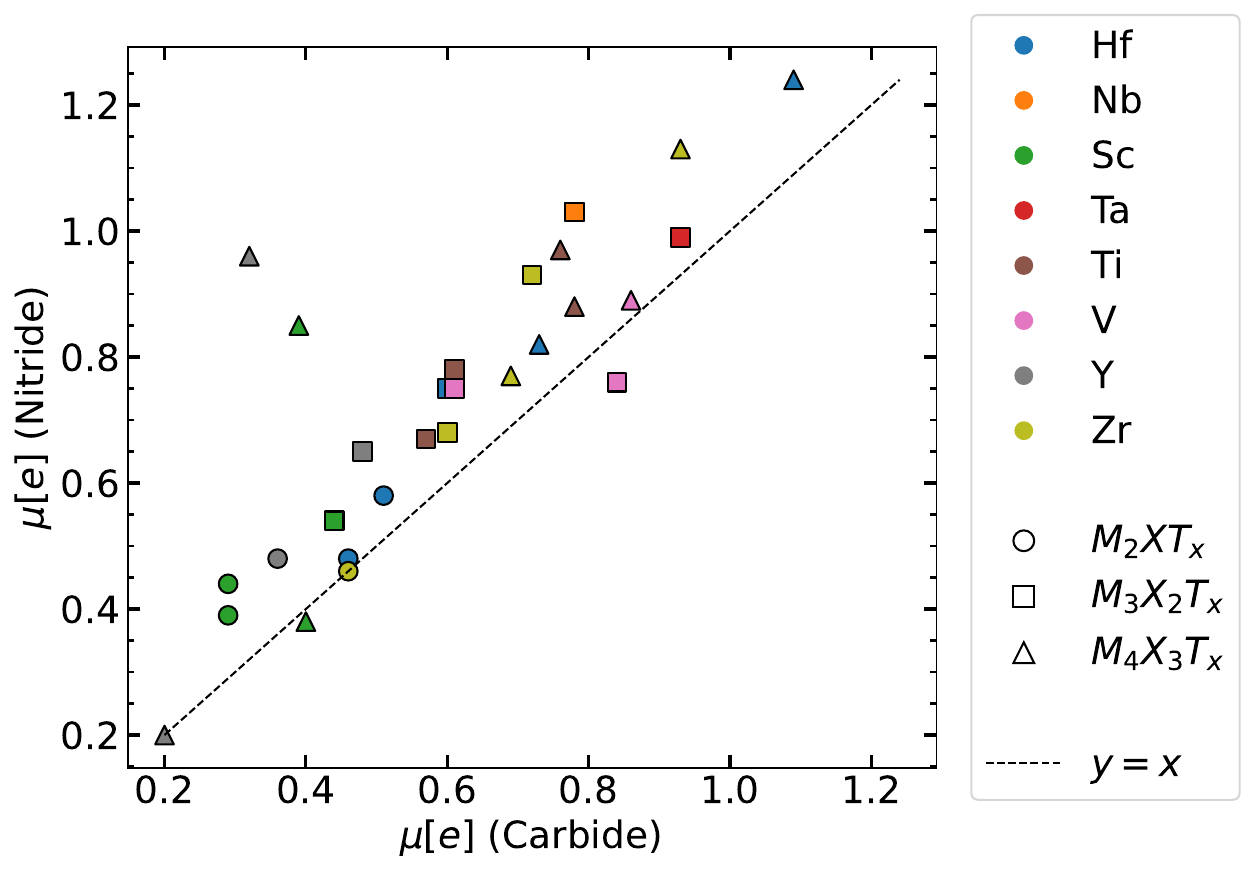}	
		\caption{Flexoelectric coefficient $\mu$ for the MXene monolayers, averaged over the armchair and zigzag directions,  with  $X$ being either carbon or nitrogen.}
		 \label{Fig:CN}
\end{figure}

In Fig.~\ref{Fig:CN}, we show the dependence of the flexoelectric coefficient on the choice of the $X$ atom, i.e., carbon or nitrogen. For a given transition metal $M$ and surface termination $T_{\rm x}$, nitride-based MXenes consistently exhibit higher flexoelectric coefficients than their carbide counterparts. Notably, $M$--C bonds are generally stronger than $M$--N bonds due to better orbital overlap, as carbon possesses more localized $p$-orbitals \cite{magnuson2018chemical, naslund2021chemical}. This suggests an inverse correlation between bond strength and flexoelectric response, consistent with the observed transition metal trend, where smaller atoms form stronger $M$--$X$ bonds and yield lower flexoelectric coefficients.

Unlike the dependence on the transition metal $M$ and  the identity of $X$ (carbon or nitrogen), the effect of surface functionalization $T_x$ on the flexoelectric coefficient is less pronounced. In particular, the ordering of flexoelectric coefficient with respect to $T_{\rm x}$ varies for the different metals. For example, in Hf-based MXenes, the trend is O $<$ OH $<$ F, whereas in Ti-based MXenes, the order is OH $<$ O $<$ F. This variability suggests that the influence of surface termination is more complex, involving subtle modifications to the local bonding environment and/or charge redistribution.

\begin{figure}[h!]
	\centering
		\includegraphics[width=0.85\textwidth]{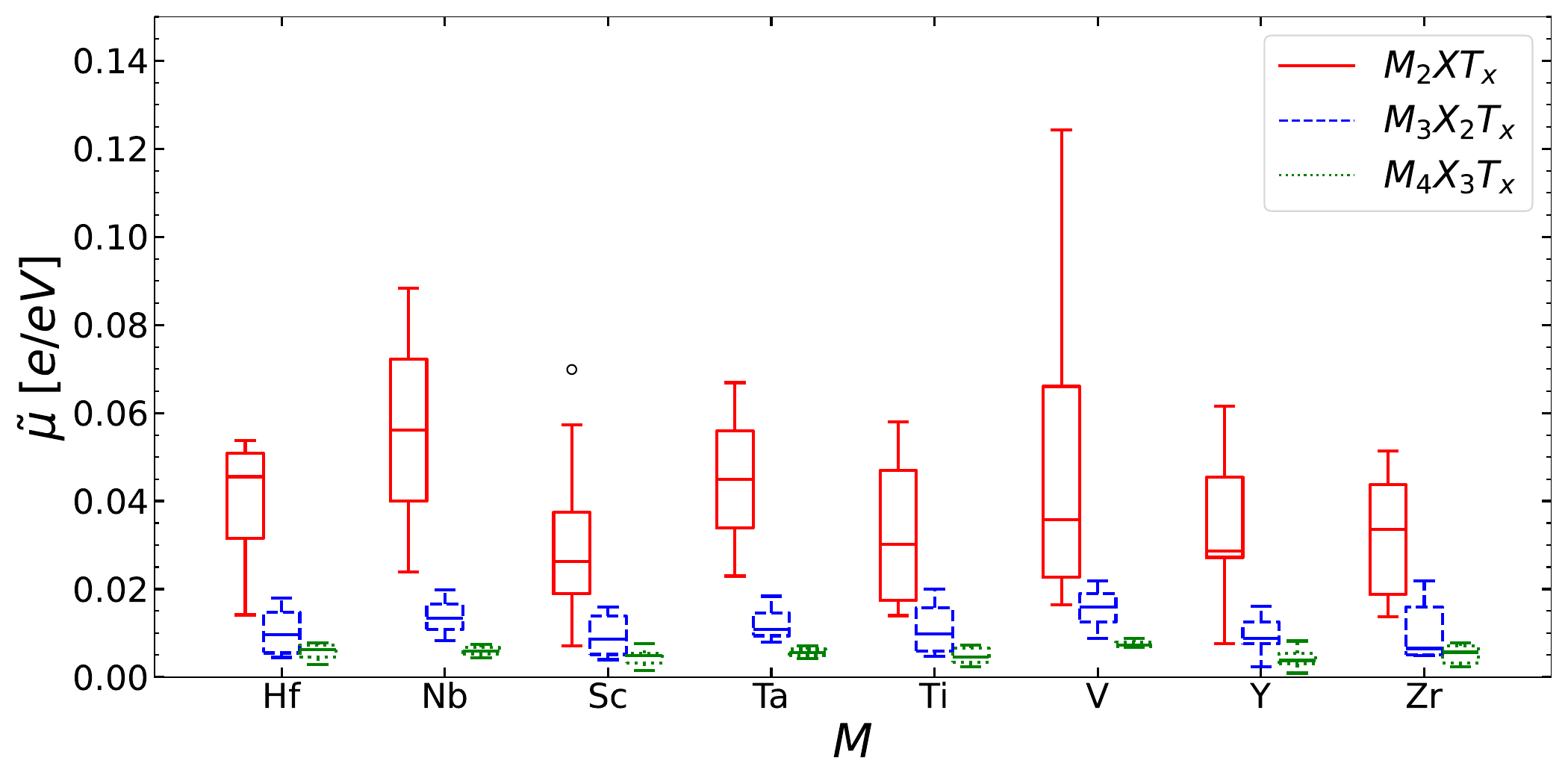}	
		\caption{Bending modulus-normalized flexoelectric coefficient $\tilde{\mu}$ for the MXene monolayers, averaged over the armchair and zigzag directions, grouped by the transition metal $M$.}
		 \label{Fig:normalizedflexo}
\end{figure}

In certain applications where bending deformations play a significant role, including energy harvesting and electromechanical sensing, the normalized flexoelectric coefficient $\tilde{\mu}$ that relates polarization to the bending moment can be more informative \cite{kumar2021flexoelectricity}.  In particular, the two flexoelectric coefficients are related as follows:
\begin{equation}
\tilde{\mu} = \frac{\mu}{D} \,,
\end{equation}
where $D$ is the bending modulus of the monolayer. The framework described above for the calculation of the flexoelectric coefficient $\mu$ also provides access to the bending modulus $D$, without the need for any additional Kohn-Sham DFT simulations \cite{kumar2020bending, Kumar2023Onthe}. The results so obtained for the bending modulus of the MXene monolayers, which are numerically accurate to within 0.1 eV, are presented in Table~\ref{tab:data_flexo_bending}. We observe that the bending modulus is nearly the same in the armchair and zigzag directions, suggesting that it is isotropic, consistent with previous Kohn-Sham DFT results in the literature for 2D materials with honeycomb structure \cite{kumar2020bending}. This translates to the coefficient $\tilde{\mu}$ also being nearly isotropic.  In Fig.~\ref{Fig:normalizedflexo}, we present the values for $\tilde{\mu}$ grouped by the transition metal $M$ in the MXene monolayers.  The normalized coefficient values span an extremely  wide range, from $\tilde{\mu}= 0.001\,e$/eV for Y$_4$N$_3$O$_2$ to $\tilde{\mu} = 0.14\,e$/eV for V$_2$N.  In comparison, the normalized coefficient of graphene from Kohn-Sham DFT calculations has been reported to be $\tilde{\mu} = 0.15\,e$/eV \cite{kumar2021flexoelectricity}. Note that there is no clear trend with respect to the transition metal $M$. In addition, the trend with respect to thickness is reversed: $M_2XT_x > M_3X_2T_x > M_4X_3T_x$. This inversion arises because thicker MXenes, although exhibiting larger  flexoelectric coefficients $\mu$, also have stiffer bending moduli $D$, which reduces their normalized response.  Also, $\tilde{\mu}$ does not display any clear trends with respect to $X$ or $T_x$.

Overall, the results presented here demonstrate the significant tunability of the flexoelectric effect in MXene monolayers through chemical composition, highlighting a rich design space for applications that leverage this electromechanical coupling.

%%%%%%%%%%%%%%%%%%%%%%%%%%%%%%%%%%%%%%%%%%%%%%
%%%%%%%%%%%%%%%%%%%%%%%%%%%%%%%%%%%%%%%%%%%%%%
%%%%%%%%%%%%%%%%%%%%%%%%%%%%%%%%%%%%%%%%%%%%%%

\section{Concluding remarks} \label{Sec:Conclusions}
We have conducted a first-principles study of flexoelectricity in MXene monolayers. In particular, we have calculated the transverse flexoelectric coefficients of 126 MXene monolayers along their two principal directions using Kohn-Sham DFT. The results revealed a wide range of flexoelectric responses, with coefficients varying from 0.19\,$e$ to 1.3\,$e$, and demonstrated near isotropy with respect to the bending direction. The choice of transition metal was found to strongly affect the flexoelectric behavior, with nitride-based MXenes consistently yielding larger coefficients than their carbide analogues. In addition, the flexoelectric coefficients generally increased with structural thickness; however, when normalized by the bending modulus, which was also computed for all 126 monolayers, this trend reversed.

Overall, this work provides a valuable reference for the transverse flexoelectric coefficients of MXene monolayers, which can help inform the design of nanoelectromechanical devices. Bilayer and multilayer systems, particularly heterostructures, have the potential to exhibit significantly enhanced flexoelectric responses and are therefore worthy of future investigation.

%%%%%%%%%%%%%%%%%%%%%%%%%%%%%%%%%%%%%%%%%%%%%%
%%%%%%%%%%%%%%%%%%%%%%%%%%%%%%%%%%%%%%%%%%%%%%
%%%%%%%%%%%%%%%%%%%%%%%%%%%%%%%%%%%%%%%%%%%%%%

\section*{Acknowledgements} 
The authors gratefully acknowledge the support of the Clifford and William Greene, Jr Professorship. This research was also supported by the supercomputing infrastructure provided by Partnership for an Advanced Computing Environment (PACE) through its Hive (U.S. National Science Foundation through grant MRI-1828187) and Phoenix clusters at Georgia Institute of Technology, Atlanta, Georgia. \vspace{-1mm}
%

%%%%%%%%%%%%%%%%%%%%%%%%%%%%%%%%%%%%%%%%%%%

%\bibliographystyle{unsrt}
\bibliography{Manuscript.bib}

%%%%%%%%%%%%%%%%%%%%%%%%%%%%%%%%%%%%%%%%%%%%

\end{document}